\documentclass[9pt,twocolumn,twoside]{osajnl}

\journal{ol} 

\setboolean{shortarticle}{true}

\usepackage{multirow}
\definecolor{red}{rgb}{0.9, 0.1, 0.1}

\definecolor{orange}{rgb}{1, 0.5, 0.0}

\usepackage{lineno}
\usepackage{float}
\usepackage{hyperref}

\usepackage[nolist]{acronym}

\title{Observing the optical modes of parametric instability}

\begin{abstract}
\ac{PI} is a phenomenon that results from resonant interactions between optical and acoustic modes of a laser cavity. This is problematic in gravitational wave interferometers where the high intra-cavity power and low mechanical loss mirror suspension systems create an environment where three mode \ac{PI} will occur without intervention.
We demonstrate a technique for real time imaging of the amplitude and phase of the optical modes of \ac{PI} yielding the first ever images of this phenomenon which could form part of active control strategies for future detectors.
\end{abstract}

\author[1,3]{Mitchell Schiworski}
\author[2,3]{Vladimir Bossilkov}
\author[2,3]{Carl Blair}
\author[1,3,*]{Daniel Brown}
\author[2,3]{Aaron Jones}
\author[1,3]{David Ottaway}
\author[2,3]{Chunnong Zhao}

\affil[1]{School of Physical Sciences, University of Adelaide, Adelaide, 5005, Australia}
\affil[2]{School of Physics, The University of Western Australia, Crawley, WA, 6009, Australia}
\affil[3]{Ozgrav, Australian Research Council Centre of Excellence for Gravitational Wave Discovery, Australia}

\affil[*]{Corresponding author: daniel.d.brown@adelaide.edu.au}

\setboolean{displaycopyright}{true}

\begin{document}

\begin{acronym}
    \acro{PI}{Parametric Instability}
    \acro{HOPTF}{High Optical Power Test Facility \cite{zhao_gingin_2006}}
    \acro{VCO}{Voltage Controlled Oscillator}
    \acro{ROI}{Region of Interest}
    \acro{ROC}{Radius of Curvature}
    \acro{aLIGO}{Advanced LIGO \cite{aasi_advanced_2015}}
    \acro{EOM}{Electro-Optic Modulator}
    \acro{QPD}{Quadrant Photo Diode}
    \acro{PBS}{Polarising Beam-Splitter}
    \acro{HG}{Hermite-Gauss}
    \acro{AMDs}{Acoustic Mode Dampers}
\end{acronym}
\newcommand{\ModeFreq}{354.8\,kHz}
 
\maketitle

\section{Introduction}

The current generation of gravitational wave interferometers operate with $\approx100$ kW of intra-cavity optical power \cite{PhysRevLett.116.061102}. Such high powers are a fundamental necessity to reduce quantum shot noise to levels that allow the detection of gravitational waves. Future detector designs require intracavity optical powers in excess of 1 MW to further reduce quantum shot noise and will be critically important for high-frequency detectors~\cite{ackley_neutron_2020}. Also limiting these detectors are thermal noises in the test masses which is reduced by using high-quality factor materials, such as fused silica. This combination of features leads to a problematic phenomenon known as \textit{three mode parametric instabilities} (PI)~\cite{BRAGINSKY2001331}. This process is an opto-mechanical interaction between two cavity optical modes and an acoustic mode of the mirror. Surface vibrations (resulting from thermal or radiation pressure affects) cause Brillouin scattering of photons from the fundamental cavity mode into a higher order transverse mode of the cavity. The two optical fields beat together to produce a time-varying radiation pressure force which then further excites the surface vibration of the mirror. This feedback process can become unstable and result in the exponential transfer of power from the main cavity mode into the higher order optical mode and mirror acoustic mode~\cite{EVANS2010665} eventually leading to a loss of control and operability of the detector.

\ac{aLIGO} was the first long-baseline gravitational wave detector to be affected by PI~\cite{evans_observation_2015} and has employed several mitigation schemes. The resonance frequency of the problematic higher order transverse optical modes can be adjusted by thermally deforming the mirror surface curvature \cite{zhao_parametric_2005}, reducing the gain in the PI feedback loop. As the stored optical powers in detectors increase the thermal actuators need to also compensate for thermally induced deformations limiting their ability to simultaneously suppress PI. It was then that the electrostatic drives were used directly to extract energy from the mechanical mode~\cite{blair_first_2017}. Around 40 independent control loops were required to control individual instabilities, rendering this method untenable once power was further increased. Dynamic thermal compensation~\cite{hardwick_demonstration_2020} has also been shown to suppress transient instabilities. Finally, \ac{AMDs} were installed on the test masses to extract energy from the mechanical resonator \cite{biscans_suppressing_2019}. \ac{AMDs} are electro-mechanical tuned mass dampers that introduce a frequency dependent mechanical loss between 15 and 80\,kHz in each test mass.
Over the last few years \ac{aLIGO}, Advanced VIRGO~\cite{acernese_advanced_2014}, and KAGRA~\cite{akutsu_overview_2020} have operated with intra-cavity power of 240\,kW~\cite{buikema_sensitivity_2020}, 130\,kW~\cite{galaxies8040085} and 1\,kW~\cite{KAGRAStatus_2020} respectively, while they are designed to operate at 800kW, 650\,kW and 400\,kW. It is so far not clear that current \ac{PI} mitigation strategies will be adequate in the future as higher optical powers are used.

An alternative \ac{PI} control strategy is to suppress the problematic transverse optical mode resonant in the cavity. Suppression has been demonstrated with feedback control by injecting a transverse mode with the opposite phase \cite{vlad_paper,zhang_enhancement_2010}. The scheme is attractive as the complexity is reduced when compared to control schemes that act on individual mechanical modes as there are $\sim$10 times fewer problematic optical modes compared to problematic mechanical resonances in gravitational wave detectors.
However, such a scheme necessitates a detailed knowledge of the optical modes produced by \ac{PI} to determine what modes to inject back in and with what phase. Previously work has been limited to information provided by the four outputs of a \ac{QPD} \cite{blair_first_2017,blair_parametric_2017}. Optical mode feedback suppression was demonstrated by sensing the beat between the fundamental and higher order transverse mode in the cavity in transmission of the cavity~\cite{vlad_paper}. The signal was then used to generate a sideband at the required transverse mode frequency. There was enough spurious coupling of the input beam to the higher order optical mode to allow suppression of the transverse mode and parametric instability. In this paper we demonstrate for the first time the direct imaging of the transverse optical mode generated by \ac{PI} providing greater spatial detail than possible with a \ac{QPD}.

\begin{figure}[h]
    \centering
    \includegraphics[width=\linewidth]{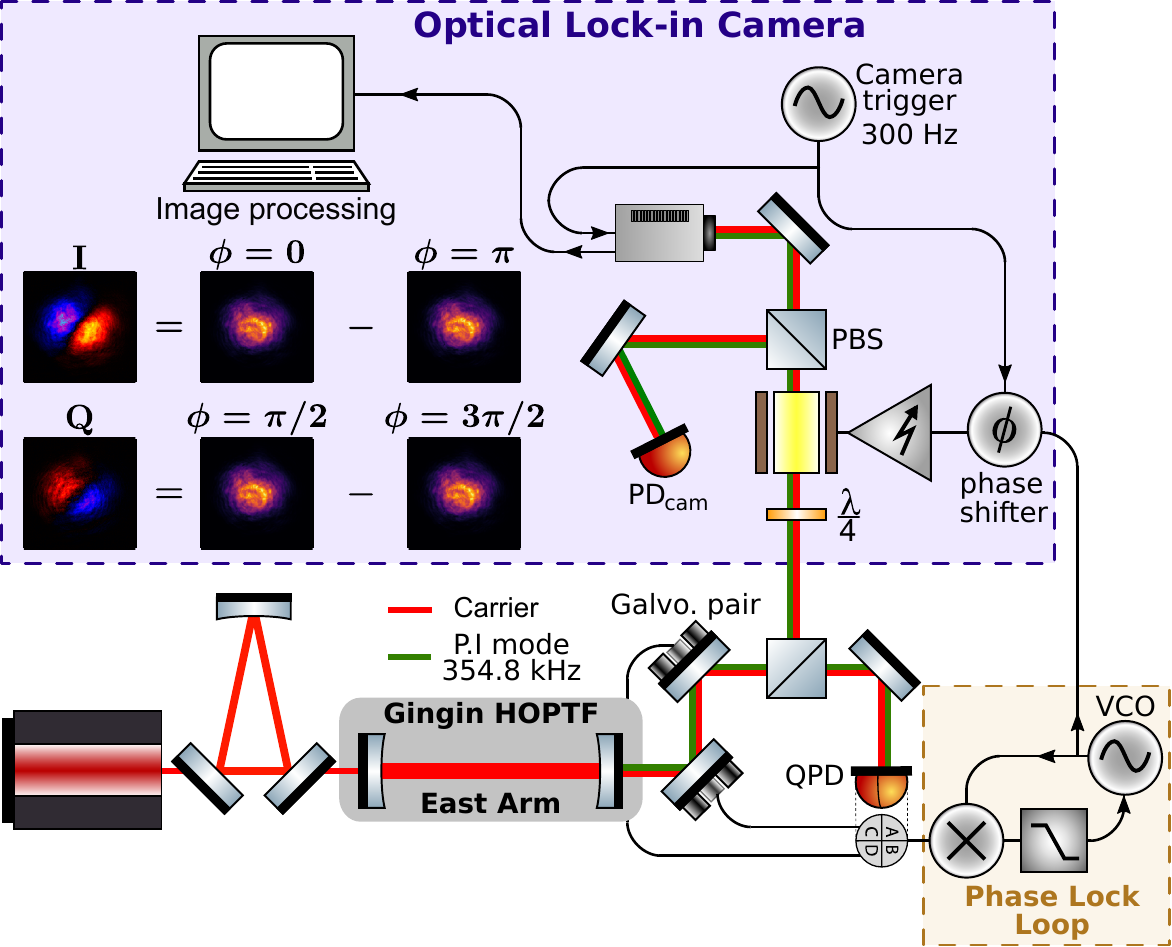}
    \caption{Schematic of the experiment. A \ModeFreq \ac{PI} sideband is intentionally excited inside the Gingin \ac{HOPTF} east arm cavity. The transmitted cavity beam is stabilized in pitch \& yaw with a pair of galvanometers and a \ac{QPD}. The pitch signal from the \ac{QPD} is used to phase lock a \ac{VCO} to the beat between the carrier and \ac{PI} sideband. The phase-shifted \& amplified \ac{VCO} signal drives the amplitude modulator component of the optical lock-in camera. The phase of the modulation is incremented by $\pi/2$ synchronously as the camera is triggered at 300 Hz. Each set of 4 images taken is digitally subtracted as shown to create amplitude and phase maps of the beat field at a rate of 75 Hz. The PD$_\text{cam}$ photodiode is not necessary but practically useful for monitoring the amount of optical modulation achieved.}
    \label{fig:experiment_schematic}
\end{figure}

Our technique uses an optical lock-in camera \cite{cao_optical_2020} to image the transverse amplitude and phase of the beat between the main cavity field and the sideband field generated by \ac{PI}. We demonstrated this experimentally at the Gingin \ac{HOPTF} facility. The first observation of three mode \ac{PI} occurred at the \ac{HOPTF} \cite{zhao_parametric_2015} and since then much of the \ac{PI} mitigation strategies developed for the gravitational wave community have also been demonstrated at this facility. Fig. \ref{fig:experiment_schematic} shows a schematic of the experiment layout.

The \ac{HOPTF} East Arm has a 74\,m suspended optical cavity designed to mimic dynamics and in particular the opto-mechanical dynamics of gravitational wave detectors. The cavity has a finesse of 14,000~\cite{fang_revealing_2021} and the optical power buildup in the cavity can be as high as $\sim$\,30\,kW.  The fused silica end mirrors have a diameter of 10\,cm, thickness of 5\,cm and weight of 880\,g.
The suspended cavity is illuminated by a single frequency 50\,W fiber laser, however the cavity optical power is limited by opto-mechanical angular instabilities~\cite{liu_angular_2018}.

The cavity is near concentric with a g-factor product that can be tuned between 0.98 and 0.7 with a combination of thermal actuation and mirror aberration that results in a decrease in \ac{ROC} towards the edge of the mirrors. This large g-factor tuning range allows many parametric instabilities to be studied between 100-400\,kHz. In this report we focus on an instability that was observed at \ModeFreq. This is observed when the beam position is far from center on the end mirror, beyond a position with accurate figure error maps and therefore the cavity g-factor is not well known. The reason for choosing this instability is that it can be sustained for long periods of time~\cite{liu_optomechanical_2019} and allows the cavity to stay locked while the mode amplitude reaches saturation levels as predicted by Danilishin et. al.~\cite{danilishin_time_2014}.

\begin{figure*}[ht]
    \centering
    \includegraphics[width=\textwidth]{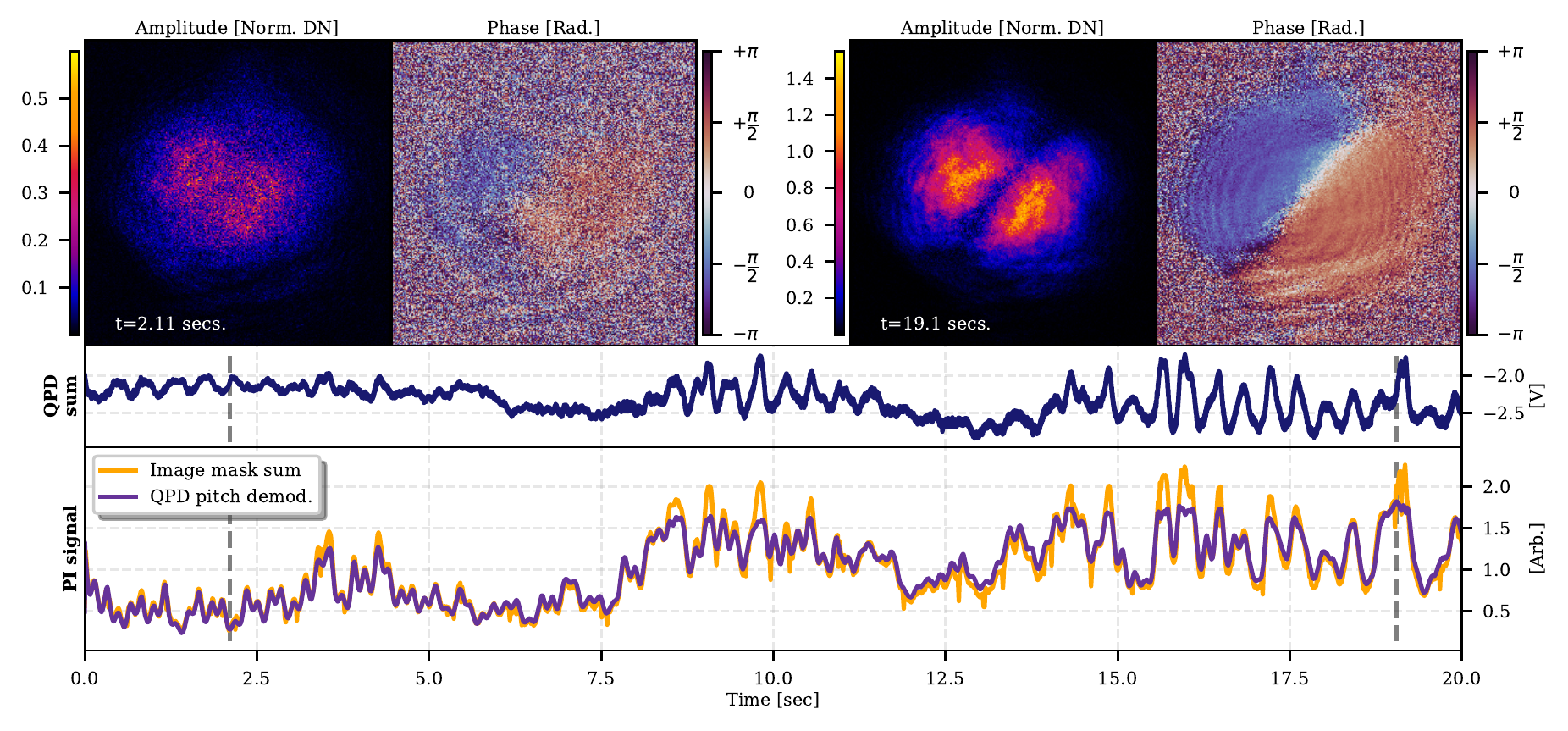}
    \caption{Amplitude \& phase of the beat field as \ac{PI} is excited inside the cavity. From top to bottom: Amplitude \& phase maps of the beat field measured with the optical lock-in camera at low (left) and high (right) levels of \ac{PI}. Transmitted beam power, measured from the \ac{QPD} sum channel. Amplitude of the \ac{PI} mode, measured via digitally demodulating the \ac{QPD} pitch channel overlayed with sum of the image counts using a \ac{HG}$_{10}$ mask (mask is shown Fig. \ref{fig:noise_floor_scatter}).}
    \label{fig:double_column_images}
\end{figure*}

A detailed description of the optical lock-in camera can be found in \cite{cao_optical_2020}. The operating principle is analogous to demodulation with a photodiode, where information of particular spectral components in a signal are recovered via mixing and low-passing the signal with a local oscillator at the frequency of interest. Here the mixing is performed optically by amplitude modulating the incident optical field, and the modulated field is imaged with a camera which also serves as a low pass filter. Each camera pixel then behaves as a single demodulated photodiode. Images taken sequentially at modulation phases of $\phi=0,\pi$ and $\phi=\pi/2,3\pi/2$ are subtracted from each other to create images of the in-phase (\textbf{I}) and in-quadrature (\textbf{Q}) components of the optical field at the frequency of interest. A sCMOS Andor Zyla 4.2 camera was configured to capture images at a 256x256 resolution. Images were taken at 300 Hz, resulting in pairs of \textbf{I} and \textbf{Q} images at a quarter of that frame rate. 

\begin{figure}[h]
    \centering
    \includegraphics[width=\linewidth]{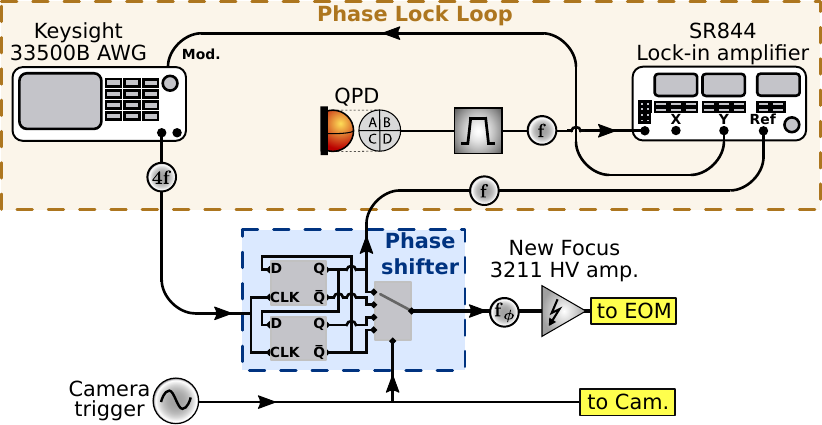}
    \caption{A diagram of the phase lock loop and phase shifter circuit. The \ac{QPD} measures the \ac{PI} signal at frequency \textbf{f}, a \ac{VCO} generates a signal at \textbf{4f} which is input to the phase shifter circuit. This circuit consists of two D-type flipflops which down-converts the signal to frequency \textbf{f} at four different phase quadratures which are sent to a 4-way switch. One of these is picked off to complete the phase lock loop, where the Mod. input is used to correct the frequency of the \ac{VCO} signal. The switch is controlled by the camera trigger such that the output of the switch \textbf{f}$_{\phi}$ is locked to the PI signal \textbf{f} but has a phase delay that increments by $\pi/2$ each camera trigger. This signal is then amplified to $\pm 200\text{ V}$ with a high voltage amplifier.}
    \label{fig:phase_lock_loop}
\end{figure}

Fluctuations in beam intensity or position results in imperfect image subtraction that causes noise artifacts in the \textbf{I} \& \textbf{Q} images. To combat beam motion caused by the suspended cavity mirrors, a PID control loop using a \ac{QPD} and pair of galvanometers was used to stabilize low frequency spot motions on the camera. As the cavity is close to concentric, the beam motion is dominated by angular motion in the cavity; controlling this motion was found to be acceptable to stop spot motion on the camera. The intensity of the transmitted beam was not stabilized and fluctuated depending on the cavity alignment, as such it was necessary to normalise the images before subtraction during post-processing. 

The optical lock-in camera requires a stable local oscillator signal to drive the Pockels cell that is matched to the phase and frequency relationship between the \ac{PI} mode and the carrier. To generate the local oscillator we used a lock-in amplifier to extract the beat on transmission of the arm using a QPD to break the modal orthogonality between the carrier and \ac{PI} sideband. 
Fig. \ref{fig:phase_lock_loop} shows a diagram of the electronics used to shift the phase of the local oscillator in $\pi/2$ increments.

The amplitude and phase profiles of the \ac{PI} mode are shown in Fig. \ref{fig:double_column_images}, these are constructed from the \textbf{I} \& \textbf{Q} images with $(\textbf{I}^2+\textbf{Q}^2)^{-1/2}$ and $\tan^{-1} \left[ \textbf{Q}/\textbf{I} \right]$ respectively.
These images show that the optical mode responsible for this \ac{PI} is a \ac{HG}$_{10}$ mode. Least-squares fits show that the mode axis is rotated by $\approx 41$ degrees.
As the \ac{HOPTF} has a shorter arm length compared to detectors like LIGO, we expect the optical \ac{PI} modes will be of lower order. This is because the optical transverse mode frequencies will be more likely to align with that of the mirror acoustic modes. In kilometer scale interferometers a mix of higher order modes will contribute to \ac{PI}, as seen in aLIGO~\cite{evans_observation_2015}.

\begin{figure}[h]
    \centering
    \includegraphics[width=\linewidth]{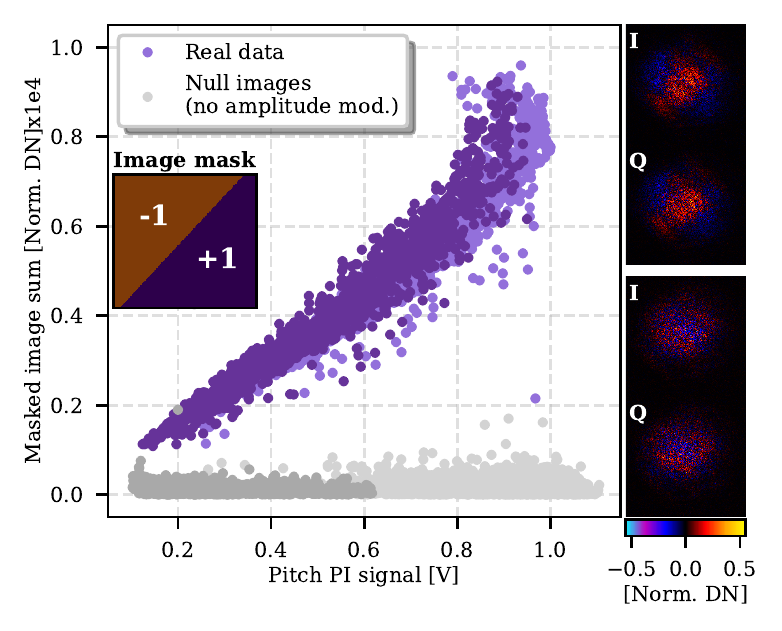}
    \caption{Masked image sum compared against the demodulated \ac{QPD} pitch signal. Sets of null images are taken separately when no amplitude modulation is applied to the field incident on the camera. Two examples of these null images are shown on the right. These images are indicative of the measurement noise as in this case the subtraction of the images at each demodulation phase should result in no signal. The lighter shades show separate measurements taken at different times.}
    \label{fig:noise_floor_scatter}
\end{figure}

Also shown in Fig. \ref{fig:double_column_images} are results from a 20 second data capture as \ac{PI} is excited within the cavity. The masked sum of the optical lock-in camera images is overlayed with the demodulated \ac{QPD} pitch signal to show the validity of the measurement. The mask parameters were extracted from a \ac{HG}$_{10}$ least-squares fit applied to the images, this mask is shown in Fig. \ref{fig:noise_floor_scatter}. From the figure it can be seen that the two measurements closely agree, apart from where electronic saturation occurs in the \ac{QPD} measurements and for glitch type images. The latter can occur when the phase lock slips and affects the optical demodulation causing a lower observed signal in the images compared to the \ac{QPD}, or when significant beam motion or intensity fluctuations occur between frames resulting in imperfect subtraction and a conversely higher observed signal in the images compared to the \ac{QPD}. The occurrence of these glitches is related to the general stability of the cavity, which suffers at high amplitude \ac{PI} as the control signals become polluted. This is seen in the sum channel of the \ac{QPD}, where after $\text{t}\approx7.5$ secs the transmitted power begins to drastically fluctuate.

A direct comparison of the demodulated \ac{QPD} pitch signal and masked image sums is shown in Fig. \ref{fig:noise_floor_scatter}. In order to gain an estimate on the noise floor of the optical lock-in camera measurements, separate sets of null images were taken. These images are taken without applying voltage to the Pockels cell, resulting in no amplitude modulation of the field incident on the camera. In this case subsequent images taken by the camera should be identical and hence perfectly subtract leaving no signal in the \textbf{I} and \textbf{Q} images. Two examples of these reference images are shown in Fig. \ref{fig:noise_floor_scatter}. Photon shot noise fundamentally limits the floor of this subtraction \cite{cao_optical_2020}, however in this experiment intensity fluctuations and movement of the beam on the camera between frames is the dominating noise source.

The comparatively large levels of \ac{PI} observed here allowed for high signal-to-noise ratio images of the sideband field. However in \ac{aLIGO}, the observed levels of \ac{PI} before the loss of lock and detector function are orders of magnitude lower and imaging these would require greater sensitivities than have yet been achieved with this optical lock-in camera.

In \ac{aLIGO} the amplitude of the mirror mechanical mode will reach $a\sim3\times10^{-12}m$ (RMS) before electronic saturation~\cite{evans_observation_2015}. Assuming a mode overlap $\beta=10\%$ and an optical gain of $G\sim130$ the relative amplitude of the sideband to the carrier is $E_{sb}/E_{c}\sim \frac{\pi}{\lambda}a\beta G \approx1.2\times10^{-4}$.
Following the notation of \cite{cao_optical_2020}, imaging this sideband field would require a sensitivity of $10\log _{10}\left(|E_{sb}|^2/|E_{c}|^2\right) = -78\,\text{dBc}$
The single pixel sensitivity demonstrated in \cite{cao_optical_2020} of this type of optical lock-in camera was $-62$\,dBc, limited by uncontrolled laser intensity noise. The camera is ultimately limited by its dynamic range and readout noise (~86 dBc for the Zyla 4.2) and photo-electron shot noise. For a Gaussian beam with radius $w$ incident on an $N\times N$ sensor of pixels with area $A$ and photo-electron well depth $W_d$ the total number of photo-electrons is $N_e \approx \pi w^2 W_d (2A)^{-1}$ (assuming the central pixel is fully illuminated to fill $W_d$ during the exposure time and the beam is smaller than the sensor $w \ll N\sqrt{A}$). The shot-noise limited sensitivity for a beam illuminating a full frame would then be $10\log _{10}\left([\sqrt{2 N_e}/(2 N_e)]^2\right) \approx -100$dBc using $w=4$mm, $W_d=30000$, $A=(6.5\textrm{um})^2$, $N=2048$.

Realisation of this technique within \ac{aLIGO} or other detectors would require increasing the $E_{sb}/E_{c}$ ratio demonstrated in this work. This could be achieved with better control over beam intensity and beam movements, which is already present in operating detectors. Alternative measurement locations might also be preferable to transmission of the cavities, such as at the dark port of the Michelson which rejects a large fraction of $|E_c|$. Other heterodyne imaging techniques, such as the mechanically scanning phase camera \cite{agatsuma_phase_2016} could also be explored.

In this work we have demonstrated a novel technique for directly imaging the optical modes of \ac{PI}. This technique provides high resolution images of the transverse amplitude and phase of the optical modes. This will allow the sensing of much higher order modes than is possible with a \ac{QPD} and for modal analysis to be used to identify instabilities resulting from mixtures of optical modes. In gravitational wave interferometers such as \ac{aLIGO}, \ac{PI} is an unwanted effect that can significantly reduce sensitivity. This will become a more significant issue for future detectors whose designs require higher circulating intracavity power. As such the study of the optical modes responsible for \ac{PI} inside these detectors is necessary for the design of systems to suppress them. We have demonstrated this technique at the Gingin \ac{HOPTF}, however higher sensitivities need to be achieved before this technique is capable of imaging \ac{PI} within the \ac{aLIGO} detector.

\section*{Funding}
This project was funded by the Australian Research Council grant CE170100004.  CB is funded by ARC DE190100437.

\section*{Disclosures}
The authors declare no conflicts of interest.

\bibliography{bib_file}

\begin{thebibliography}{10}
\newcommand{\enquote}[1]{``#1''}

\bibitem{PhysRevLett.116.061102}
{LIGO Scientific Collaboration} and {Virgo Collaboration},
  {\protect\JournalTitle{Phys. Rev. Lett.}} \textbf{116}, 061102 (2016).

\bibitem{ackley_neutron_2020}
K.~Ackley \emph{et~al.}, {\protect\JournalTitle{Publications of the
  Astronomical Society of Australia}} \textbf{37} (2020). Publisher: Cambridge
  University Press.

\bibitem{BRAGINSKY2001331}
V.~Braginsky, S.~Strigin, and S.~Vyatchanin, {\protect\JournalTitle{Physics
  Letters A}} \textbf{287}, 331 (2001).

\bibitem{EVANS2010665}
M.~Evans, L.~Barsotti, and P.~Fritschel, {\protect\JournalTitle{Physics Letters
  A}} \textbf{374}, 665 (2010).

\bibitem{aasi_advanced_2015}
T.~L.~S. Collaboration, {\protect\JournalTitle{Classical and Quantum Gravity}}
  \textbf{32}, 074001 (2015). Publisher: IOP Publishing.

\bibitem{evans_observation_2015}
M.~Evans \emph{et~al.}, {\protect\JournalTitle{Physical Review Letters}}
  \textbf{114}, 161102 (2015). Publisher: American Physical Society.

\bibitem{zhao_parametric_2005}
C.~Zhao, L.~Ju, J.~Degallaix, S.~Gras, and D.~G. Blair,
  {\protect\JournalTitle{Phys. Rev. Lett.}} \textbf{94}, 121102 (2005).
  Publisher: American Physical Society.

\bibitem{blair_first_2017}
C.~Blair \emph{et~al.}, {\protect\JournalTitle{Phys. Rev. Lett.}} \textbf{118},
  151102 (2017). Publisher: American Physical Society.

\bibitem{hardwick_demonstration_2020}
T.~Hardwick, V.~J. Hamedan, C.~Blair, A.~C. Green, and D.~Vander-Hyde,
  {\protect\JournalTitle{Classical and Quantum Gravity}} \textbf{37}, 205021
  (2020). Publisher: IOP Publishing.

\bibitem{biscans_suppressing_2019}
S.~Biscans, S.~Gras, C.~Blair, J.~Driggers, M.~Evans, P.~Fritschel,
  T.~Hardwick, and G.~Mansell, {\protect\JournalTitle{Physical Review D}}
  \textbf{100}, 122003 (2019). Publisher: American Physical Society.

\bibitem{acernese_advanced_2014}
F.~Acernese \emph{et~al.}, {\protect\JournalTitle{Classical and Quantum
  Gravity}} \textbf{32}, 024001 (2014). Publisher: IOP Publishing.

\bibitem{akutsu_overview_2020}
T.~Akutsu \emph{et~al.}, {\protect\JournalTitle{arXiv:2005.05574 [astro-ph,
  physics:gr-qc, physics:physics]}}  (2020). ArXiv: 2005.05574.

\bibitem{buikema_sensitivity_2020}
A.~Buikema, C.~Cahillane, G.~Mansell, C.~Blair \emph{et~al.},
  {\protect\JournalTitle{Physical Review D}} \textbf{102}, 062003 (2020).
  Publisher: American Physical Society.

\bibitem{galaxies8040085}
A.~Allocca, D.~Bersanetti, J.~Casanueva~Diaz, C.~De~Rossi, M.~Mantovani,
  A.~Masserot, L.~Rolland, P.~Ruggi, B.~Swinkels, E.~N. Tapia San~Martin,
  M.~Vardaro, and M.~Was, {\protect\JournalTitle{Galaxies}} \textbf{8} (2020).

\bibitem{KAGRAStatus_2020}
S.~Miyoki, \enquote{{Current status of KAGRA},} in \emph{Ground-based and
  Airborne Telescopes VIII,} , vol. 11445 H.~K. Marshall, J.~Spyromilio, and
  T.~Usuda, eds., International Society for Optics and Photonics (SPIE, 2020),
  pp. 192 -- 204.

\bibitem{vlad_paper}
V.~Bossilkov \emph{et~al.}, \enquote{Demonstration of parametric instability
  suppression through optical feedback,}  (in preparation).

\bibitem{zhang_enhancement_2010}
Z.~Zhang, C.~Zhao, L.~Ju, and D.~Blair, {\protect\JournalTitle{Phys. Rev. A}}
  \textbf{81} (2010).

\bibitem{blair_parametric_2017}
C.~D. Blair, \enquote{Parametric instability in gravitational wave detectors,}
  Doctoral {Thesis} (2017).

\bibitem{cao_optical_2020}
H.~T. Cao, D.~D. Brown, P.~J. Veitch, and D.~J. Ottaway,
  {\protect\JournalTitle{Optics Express}} \textbf{28}, 14405 (2020). Publisher:
  Optical Society of America.

\bibitem{zhao_gingin_2006}
C.~Zhao \emph{et~al.}, {\protect\JournalTitle{Journal of Physics: Conference
  Series}} \textbf{32}, 368 (2006). Publisher: IOP Publishing.

\bibitem{zhao_parametric_2015}
C.~Zhao, L.~Ju, Q.~Fang, C.~Blair, J.~Qin, D.~Blair, J.~Degallaix, and
  H.~Yamamoto, {\protect\JournalTitle{Physical Review D}} \textbf{91}, 092001
  (2015). Publisher: American Physical Society.

\bibitem{fang_revealing_2021}
Q.~Fang, C.~D. Blair, C.~D. Blair, C.~Zhao, C.~Zhao, D.~G. Blair, and D.~G.
  Blair, {\protect\JournalTitle{Optics Express}} \textbf{29}, 23902 (2021).
  Publisher: Optical Society of America.

\bibitem{liu_angular_2018}
J.~Liu, V.~Bossilkov, C.~Blair, C.~Zhao, L.~Ju, and D.~G. Blair,
  {\protect\JournalTitle{Review of Scientific Instruments}} \textbf{89}, 124503
  (2018). Publisher: American Institute of Physics.

\bibitem{liu_optomechanical_2019}
J.~Liu, \enquote{Optomechanical instabilities in laser interferometer
  gravitational wave detectors,} Doctoral {Thesis} (2019).

\bibitem{danilishin_time_2014}
S.~L. Danilishin, S.~P. Vyatchanin, D.~G. Blair, J.~Li, and C.~Zhao,
  {\protect\JournalTitle{Physical Review D}} \textbf{90}, 122008 (2014).
  Publisher: American Physical Society.

\bibitem{agatsuma_phase_2016}
K.~Agatsuma, M.~van Beuzekom, L.~van~der Schaaf, and J.~van~den Brand,
  {\protect\JournalTitle{Nuclear Instruments and Methods in Physics Research
  Section A: Accelerators, Spectrometers, Detectors and Associated Equipment}}
  \textbf{824}, 598 (2016).

\end{thebibliography}

\end{document}